\documentclass[longauth]{aa} 

\usepackage{graphicx}
\usepackage{color}
\usepackage{natbib}
\usepackage{txfonts}
\usepackage{lineno}
\usepackage{comment}
\begin{document} 


   \title{Gamma-ray flaring activity of NGC\,1275 in 2016-2017 measured by MAGIC}

%
%
\author{
MAGIC collaboration:
S.~Ansoldi\inst{1,20} \and
L.~A.~Antonelli\inst{2} \and
C.~Arcaro\inst{3} \and
D.~Baack\inst{4} \and
A.~Babi\'c\inst{5} \and
B.~Banerjee\inst{6} \and
P.~Bangale\inst{7} \and
U.~Barres de Almeida\inst{7,8} \and
J.~A.~Barrio\inst{9} \and
J.~Becerra Gonz\'alez\inst{10}
% ...
}
\institute { Universit\`a di Udine, and INFN Trieste, I-33100 Udine, Italy
\and National Institute for Astrophysics (INAF), I-00136 Rome, Italy
\and Universit\`a di Padova and INFN, I-35131 Padova, Italy
\and Technische Universit\"at Dortmund, D-44221 Dortmund, Germany
\and Croatian MAGIC Consortium: University of Rijeka, 51000 Rijeka, University of Split - FESB, 21000 Split,  University of Zagreb - FER, 10000 Zagreb, University of Osijek, 31000 Osijek and Rudjer Boskovic Institute, 10000 Zagreb, Croatia.
\and Saha Institute of Nuclear Physics, HBNI, 1/AF Bidhannagar, Salt Lake, Sector-1, Kolkata 700064, India
\and Max-Planck-Institut f\"ur Physik, D-80805 M\"unchen, Germany
\and now at Centro Brasileiro de Pesquisas F\'isicas (CBPF), 22290-180 URCA, Rio de Janeiro (RJ), Brasil
\and Unidad de Part\'iculas y Cosmolog\'ia (UPARCOS), Universidad Complutense, E-28040 Madrid, Spain
\and Inst. de Astrof\'isica de Canarias, E-38200 La Laguna, and Universidad de La Laguna, Dpto. Astrof\'isica, E-38206 La Laguna, Tenerife, Spain
\and University of \L\'od\'z, Department of Astrophysics, PL-90236 \L\'od\'z, Poland
\and Deutsches Elektronen-Synchrotron (DESY), D-15738 Zeuthen, Germany
\and ETH Zurich, CH-8093 Zurich, Switzerland
\and Institut de F\'isica d'Altes Energies (IFAE), The Barcelona Institute of Science and Technology (BIST), E-08193 Bellaterra (Barcelona), Spain
\and Universit\`a  di Siena and INFN Pisa, I-53100 Siena, Italy
\and Universit\"at W\"urzburg, D-97074 W\"urzburg, Germany
\and Finnish MAGIC Consortium: Tuorla Observatory and Finnish Centre of Astronomy with ESO (FINCA), University of Turku, Vaisalantie 20, FI-21500 Piikki\"o, Astronomy Division, University of Oulu, FIN-90014 University of Oulu, Finland
\and Departament de F\'isica, and CERES-IEEC, Universitat Aut\'onoma de Barcelona, E-08193 Bellaterra, Spain
\and Universitat de Barcelona, ICC, IEEC-UB, E-08028 Barcelona, Spain
\and Japanese MAGIC Consortium: ICRR, The University of Tokyo, 277-8582 Chiba, Japan; Department of Physics, Kyoto University, 606-8502 Kyoto, Japan; Tokai University, 259-1292 Kanagawa, Japan; The University of Tokushima, 770-8502 Tokushima, Japan
% ...
}

\date{Received ...; accepted ...}

\abstract
{
We report on the detection of flaring activity from the Fanaroff-Riley~I radio galaxy NGC\,1275 in very-high-energy (VHE, E $>$ 100 GeV) gamma rays with the MAGIC telescopes.
The observations were performed between 2016 September and 2017 February, as part of a monitoring program. The brightest outburst with $\sim1.5$ times the Crab Nebula flux above 100\,GeV (C.U.) was observed during the night between 2016 December 31 and 2017 January 1. The flux is fifty times higher than the mean flux previously measured in two observational campaigns between 2009 October and 2010 February and between 2010 August and 2011 February. Significant variability of the day-by-day light curve was measured. The shortest flux-doubling time-scales was found to be of $(611\pm101)$\,min. The spectra calculated for this period are harder and show a significant curvature with respect to the ones obtained in the previous campaigns. The combined spectrum of the MAGIC data during the strongest flare state and simultaneous data from the \textit{Fermi}-LAT around 2017 January 1 follows a power-law with an exponential cutoff at the energy $(492\pm35)$\,GeV. We further present simultaneous optical flux density measurements in the R-band obtained with the KVA telescope and investigate the correlation between the optical and gamma-ray emission.\\
Due to possible internal pair-production, the fast flux variability constrains the Doppler factor to values which are inconsistent with a large viewing angle as observed in the radio band. We investigate different scenarios for the explanation of fast gamma-ray variability, namely emission from: magnetospheric gaps, relativistic blobs propagating in the jet (mini-jets) or external cloud (or star) entering the jet. We find that the only plausible model to account for the luminosities here observed would be the production of gamma rays in a magnetospheric gap around the central black hole only in the eventuality of an enhancement of the magnetic field threading the hole from its equipartition value with the gas pressure in the accretion flow. The observed gamma-ray flare therefore challenges all the discussed models for fast variability of VHE gamma-ray emission in active galactic nuclei.
}

               
\authorrunning{M.L. Ahnen et al.}
\titlerunning{Flaring activity of NGC\,1275 in 2016-2017 measured by MAGIC}
\maketitle
%

\section{Introduction}

The majority of gamma-ray detected Active Galactic Nuclei (AGN), namely \textit{blazars} \citep{fermi2017}, are characterized by a small angle between the jet axis and the line of sight of the observer (viewing angle $\theta$). Doppler boosting of their non-thermal emission, conventionally explained as due to an emitting region moving relativistically along the jet axis, accommodates the enormous luminosities observed ($ \sim 10^{49}$ erg s$^{-1}$), along with features like fast flux variability. In this broadly accepted scenario \citep{Mastichiadis97} the gamma-ray emission is produced via Comptonization of internal or external radiation fields.
Variabilities shorter than the light crossing time at the black hole event horizon could challenge the aforementioned model: even though they can still be described by adapting extreme parameters, such as large Doppler factors \citep{begelman2008}, theoretical alternatives were formulated over the years \citep{giannios10, levinson2011, Tavecchio2014, barkov12, hirotani16a}.
Observing huge gamma-ray luminosities and fast variability in non-blazar AGN, like radio galaxies \citep{acciari09, aleksic14b}, in which a larger viewing angle ($\theta > 10^{\circ}$) can eventually cause a de-boosting of the radiation, it is therefore an intriguing phenomenon. It provides room for discussion of the aforementioned alternative models, especially in case of the most extreme phenomena observed in \citep{aleksic14b}. 
Gamma-ray observations and exploration of the variability of non-blazar AGN is crucial to provide insights into the location and physical processes behind extragalactic non-thermal emission.  \\
NGC\,1275, also known as 3C\,84, is the central galaxy of the Perseus cluster with a redshift of $z=0.0176$ \citep{falco1999}. While the optical spectrum shows strong nuclear emission lines typical of a Seyfert galaxy \citep{Humason1932, Khachikian1974}, the morphology in the radio band reveals a Fanaroff-Riley~I type \citep{Vermeulen1994, Buttiglione2010}. The viewing angle of NCG\,1275 was inferred from radio interferometers to be $\theta=30^\circ-55^\circ$ by \citet{walker1994} and $\theta=65^\circ\pm 16^\circ$ \citet{Fujita2017} on (sub)-parsec scales. The sub-parsec radio jet shows a new component (C3) which appeared about ten years ago and keeps growing in brightness as it moves downstream the jet \citep{nagai2010, nagai2012}. Flux variability has been detected in various frequency bands: in radio \citep{Dutson2014}, optical \citep{aleksic14c}, and X-rays \citep{Fukazawa2016}, although the emission is often affected by other contributions such as the host galaxy \citep{aleksic14c} or filaments \citep{fabian2011} so that the AGN emission is difficult to extract.   \\
After the early detection with the \textit{COS\,B} satellite \citep{strong1982}, NGC\,1275 was observed in high-energy (HE, $E>100$\,MeV) gamma-ray regime with \textit{Fermi}-LAT \citep{abdo2009a} and later measured with MAGIC and VERITAS in VHE \citep{aleksic12, aleksic14c, Benbow2015}. While the measurements with \textit{Fermi}-LAT yielded flux variability on time scales of $(1.51\pm0.02)$\,d \citep{brown2011}, MAGIC measurements showed marginal flux changes on monthly scales. 
A detailed analysis of \textit{Fermi}-LAT data during 2008–2017 can be found in \citet{baghmanyan2017}.
Recently flaring activity in VHE band was reported MAGIC and VERITAS \citep{2016ATel.9689, 2016ATel.9690, 2017ATel.9929, 2017ATel.9931} \newline

This paper is structured as follows: in section 2 we report on the results of the MAGIC observations during the period between 2016 September and 2017 February and on the analysis of simultaneous \textit{Fermi}-LAT data. Gamma-ray light curves are produced along with VHE spectra for different flux states. A joint MAGIC and \textit{Fermi}-LAT spectrum is shown for the night with the strongest flux. Measurements with the KVA telescope in the optical band allow us to present also an optical - gamma-ray emission correlation study. In section 3 the physics discussion, supported by the considerations in \citet{levinson2011} and \citet{hirotani16a,hirotani16b} along with some of the analytical parametrization provided in \citet{aharonian2017}, tries to identify the emission model more suitable to explain the observed gamma-ray fluxes.

%
\section{Observational results}

\subsection{MAGIC}
The observations here reported were conducted with the MAGIC (Major Atmospheric Gamma Imaging Cherenkov) telescopes \citep{aleksic12a,AleksicSoftwareUpgrade}, two 17\,m-diameter Imaging Atmospheric Cherenkov Telescopes located at the Canary island of La Palma, Spain, designed to perform gamma-ray astronomy in the energy range from 50\,GeV to 50\,TeV \citep{aleksic12a,AleksicSoftwareUpgrade}.
NGC\,1275 was observed between 2016 September and 2017 February (MJD 57637.1--57811.9) for 63 hours mostly under dark conditions as part of a monitoring program. Seven hours of data affected by non-optimal weather conditions were discarded. The analysis of the data was performed using the standard analysis chain described in \citet{AleksicSoftwareUpgrade}. The data cover the zenith distance range of $12^\circ<\mathrm{Zd}<\,50^\circ$ \footnote{low zenith angles and low night sky background levels reflect in lower energy threshold and sensitivity \citep{AleksicSoftwareUpgrade}}.\\ 
Following \citet{AleksicSoftwareUpgrade}, we consider for the spectra the following systematic errors: 11\% for the flux normalization, 15\% for the energy scale and 0.15 for the photon index. The absorption due to the extragalactic background light (EBL) is only marginal for the redshift and the calculated energy range of the spectrum of NGC\,1275. The cutoff in the VHE spectrum due to the EBL is expected to be at energies $>10$\,TeV as discussed in \citet{ahnen16a}. \\


\subsubsection{VHE Flux Variability}

\begin{figure*}
   \centering
   \includegraphics[width=16.cm]{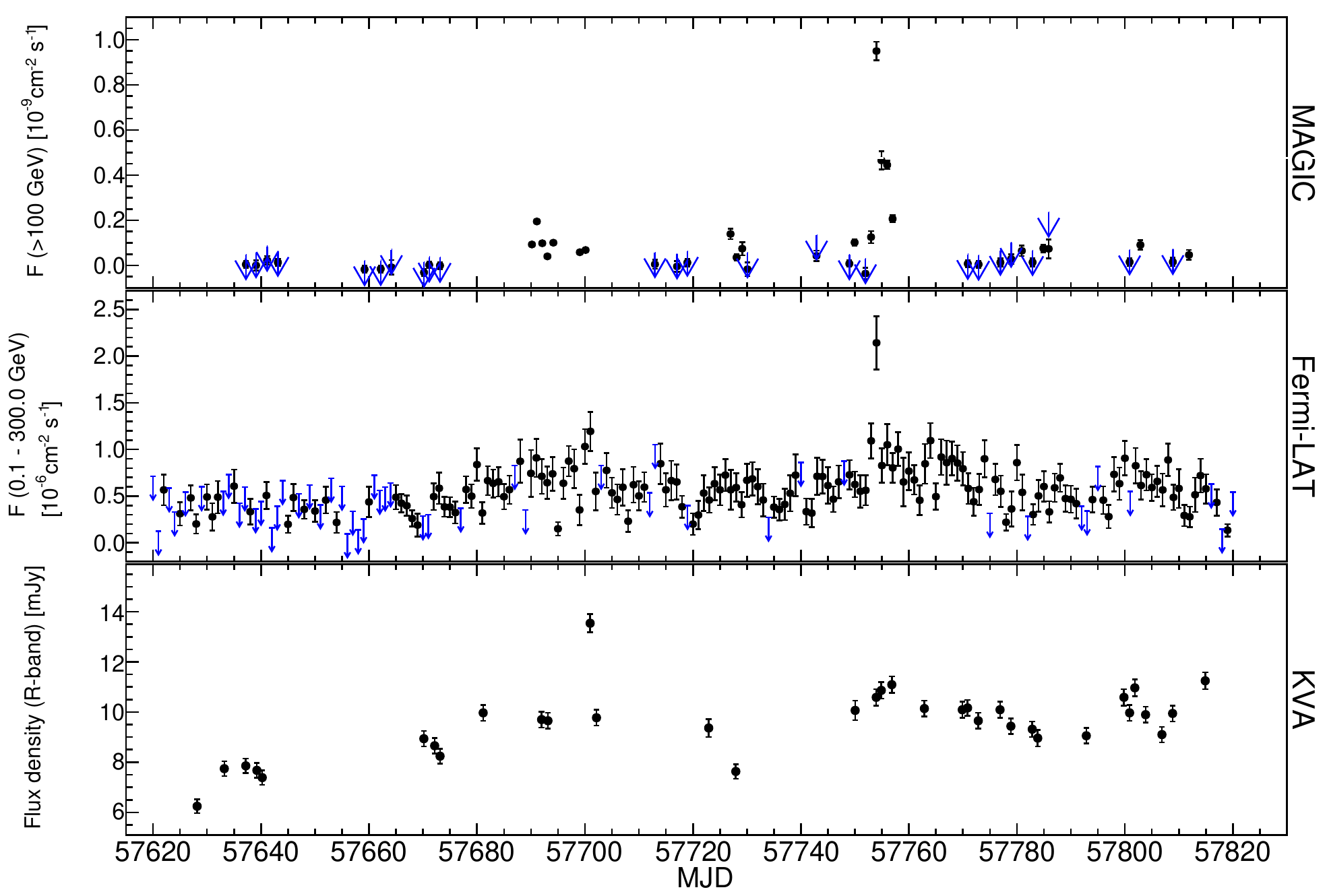}
   \caption{Light curves of NGC\,1275 between 2016 September and 2017 February in different energy bands.
   \textit{Upper panel:} Daily (black data points and blue upper limits) flux measurements above 100\,GeV obtained from MAGIC observations.  All flux upper limits are given at 95\% confidence level and were calculated assuming a total systematic uncertainty of 30\% using the \citet{rolke} method.
   \textit{Middle panel:} Daily binned fluxes (black data points) calculated from \textit{Fermi}-LAT observations in an energy range of 0.1--300.0\,GeV. Flux upper limits were estimated at 95\% confidence level in case of TS$<25$ and are shown in blue.
   \textit{Bottom panel:} R-band flux density measurements by KVA are host galaxy subtracted and corrected for galactic extinction.
  }
              \label{Fig:LC}%
    \end{figure*}

The mean flux between 2016 September to 2017 February equals $(1.19\pm0.03)\times10^{-10}$\,cm$^{-2}$\,s$^{-1}$ above 100\,GeV whereas previous measurements yielded $(1.6\pm0.3)\times10^{-11}$\,cm$^{-2}$\,s$^{-1}$ and $(1.3\pm0.2)\times10^{-11}$\,cm$^{-2}$\,s$^{-1}$ during 2009-2010 and 2010-2011, published in \citet{aleksic14c}. Thus, the mean flux reported here is seven to nine times higher.\\
The VHE daily light curve is shown in the upper panel of Fig.~\ref{Fig:LC} and is calculated assuming a power-law index of $\Gamma=3.0$. Fitting the daily light curve with a constant function yields a $\chi^2$/d.o.f. of $1574.6/44$. 
Several nights around 2016 November until 2017 January were identified with a clear high flux. We investigated the intra-night light curves for these nights but did not find strong hints for variability during the observation of typically 1-3\,h. For the night with the highest flux, 2017 January 01, 8 data runs for a total observation time of 2.4 h were collected. A fit of the run-wise LC with a constant flux returned a probability of 0.09. No individual data point deviates by more than 2 sigma from a constant fit value.

\begin{figure}
   \centering
   \includegraphics[width=8.cm]{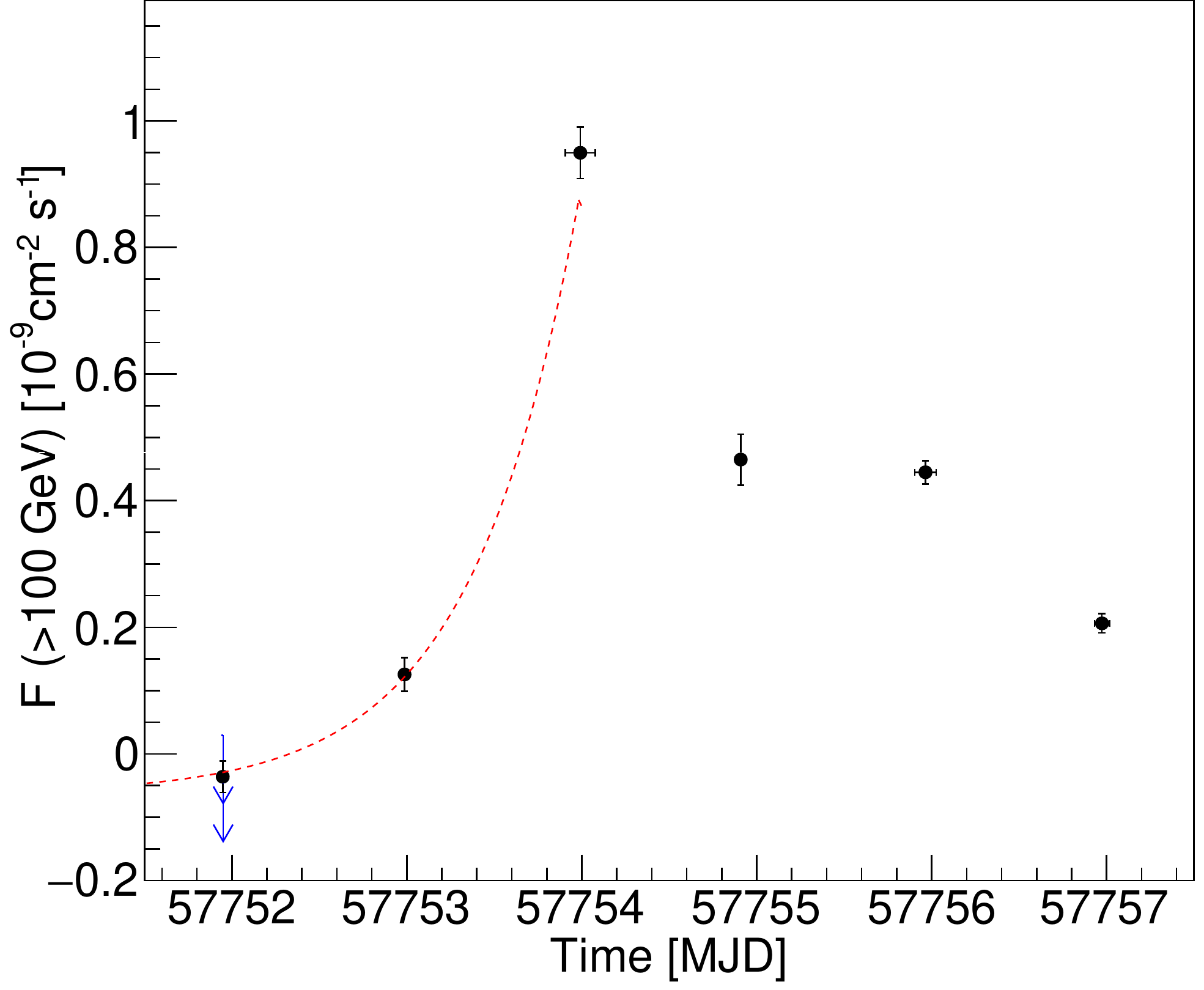}
   \caption{Zoom-in to the light curve observed by the MAGIC telescopes above 100\,GeV around 2017-01-01 together with an exponential fit. }
              \label{Fig:LC_zoom}%
    \end{figure}

In order to estimate the time-scale at which the flux has doubled, we fit the daily light curve during the brightest outburst around MJD\,57751.00--57754.02, as shown in Fig.~\ref{Fig:LC_zoom}, with the function: 
\begin{equation}
 F= F_0+F_1*2^{-\vert t-t_1\vert/\tau}
\end{equation}
where $F_0$ is the baseline flux, $F_0+F_1$ the normalization flux at the time $t_1$, and $\tau$ is the flux-doubling time-scale. 
Fixing $F_1$ to $9.5\times10^{-10}$\,cm$^{-2}$\,s$^{-1}$ and $t_1$ to MJD\,$57753.99$ yields a flux-doubling time-scale of $(611\pm101)$\,min ($\chi^2$/d.o.f. of $0.49/1$, probability of 0.49) for the rising part of the flare in 2016 December and 2017 January. Note that we only fit the measured data points, not taking into account the upper limits, the result is however consistent with them.

\subsubsection{VHE Spectral analysis}

\begin{table}
\tiny     
\caption{Parameters of the spectral fit to the VHE SED obtained with MAGIC data. The fit range is 64\,GeV--2.1\,TeV, for 2017-01-01 and for the low state and 64\,GeV--1.4\,TeV for 2017-01-02,03.
The flux normalization $f_0$ is given in units of $10^{-10}$\,TeV$^{-1}$\,cm$^{-2}$\,s$^{-1}$. 
Only statistical errors are given.}
\label{tabSpec}

power-law with exponential cutoff: $\mathrm{d}F/\mathrm{d}E=f_0\left(\frac{E}{\mathrm{300\,GeV}}\right)^{-\Gamma}\mathrm{e}^{-E/E_\mathrm{C}}$ with the cutoff energy $E_\mathrm{C}$ given in units of TeV\\

\begin{tabular}{p{17mm}p{12mm}p{12mm}p{12mm}p{7mm}p{7mm}}      
\hline\hline
Epoch        & $f_0$  	& $\Gamma$ & $E_\mathrm{C}$ &  $\chi^2$/d.o.f.	&  Prob.		  	\\
\hline
low state    & $1.14\pm0.32$    & $2.28\pm0.22$ & $0.36\pm0.11$ & 3.7/5   	&  0.60	 	\\ 
2017-01-01   & $16.1\pm2.3$     & $2.11\pm0.14$ & $0.56\pm0.11$ & 2.5/5 	&  0.78	   	\\ 
2017-01-02/03& $15.4\pm4.5$     & $1.61\pm0.25$ & $0.25\pm0.05$ & 3.81/4 	&  0.43	   	\\ 
\hline
\end{tabular}\\
\vspace{3.mm}
\\

log-parabola: $\mathrm{d}F/\mathrm{d}E=f_0\left(\frac{E}{300\,\mathrm{GeV}}\right)^{-\Gamma-\beta\log(E/300\,\mathrm{GeV})}$\\
    
\begin{tabular}{p{17mm}p{12mm}p{12mm}p{12mm}p{6mm}p{11mm}}  
\hline\hline
Epoch        & $f_0$  	& $\Gamma$  &  $\beta$ 	& $\chi^2$/	&  Prob.		  	\\
    &   	&   &  	& d.o.f.	&  		  	\\
\hline
low state    & $0.40\pm0.01$ & $3.33\pm0.04$  & $0.40\pm0.08$&30.65/5& $1.1\times10^{-5}$  	\\ 
2017-01-01   & $9.52\pm0.48$ & $2.77\pm0.05$  & $0.84\pm0.15$&7.68/5 & 0.17  		   	\\
2017-01-02/03& $4.55\pm0.29$ & $2.98\pm0.08$  & $1.37\pm0.26$&7.81/4 & 0.10		   	\\
\hline
\end{tabular}\\
\\
\end{table} 

\begin{figure}
   \centering
   \includegraphics[width=9cm]{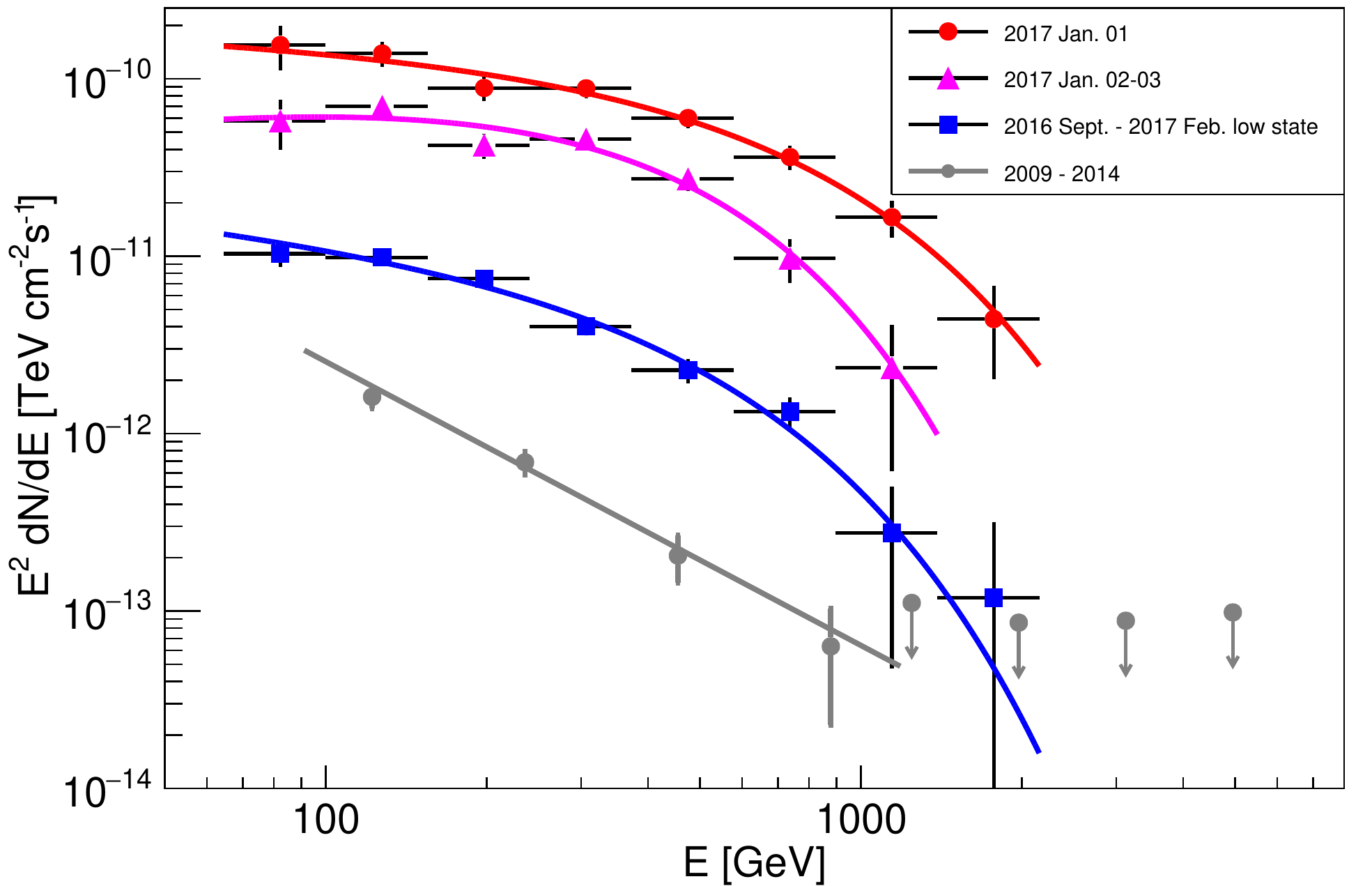}
   \caption{Measured spectral energy distributions in the VHE band during different periods. The red data points as well as the red line show the resulting SEDs from observations on 2017-01-01 while the purple points and purple line correspond to the measurement from 2017-01-02 to 2017-01-03. The SED calculated from the remaining observations between 2016 September and 2017 February is shown in blue.  All results were obtained with power-law with exponential cutoff fits. The averaged spectrum together with upper limit calculated with a photon index of $\Gamma=3.5$ from observations in 2009 to 2014 is shown in gray \citep{ahnen16a}. }
              \label{Fig:Spec}%
\end{figure} 

We divided the dataset according to the following flux states:
\begin{itemize}
\item flux $>1$\,C.U.: night of 2017-01-01 (MJD 57753.90--57754.08);
\item 0.4\,C.U. < flux < 1\,C.U.: nights of 2017-01-02 and 2017-01-03 (MJD 57754.89-57756.03); 
\item flux $<0.4$\,C.U.: remaining nights of the monitored period, noting that this low state sample also includes some weaker flares. 
\end{itemize}
For each of this dataset we calculate the spectrum. \newline
The spectral energy distributions (SEDs) are shown in Fig.~\ref{Fig:Spec} together with the long-term averaged result from observations in 2009 to 2014. Continuous line represent the result of a fit using the whole energy band while individual spectral points are calculated using the unfolding method in \citet{unfolding2007}. Due to a clear curvature, we fitted the SED with a power-law with exponential cutoff and with a log-parabola function. Parameters of the spectral fit to the VHE SED obtained with MAGIC data are given in Table~\ref{tabSpec}.\\
The power-law with exponential cutoff provides the highest probability for the fit, although the log-parabola fit can not be excluded. The cutoff energies at around 250--560\,GeV can not be a result of the EBL since its effect would start to dominate at energies above 10\,TeV. \newline
Above 1\,TeV, NGC\,1275 could still be detected with a significance of 8\,$\sigma$ using all the data and the equation derived by \citet{lima83}. Long-term observations presented in \citet{ahnen16a} for a total of 253\,h between 2009--2014 did not yield in a detection above 1\,TeV. \\

\subsection{\textit{Fermi}-LAT}
In order to further investigate the presence of a cutoff in the brightest state, to provide a more constrained spectral information, and to search for a correlation between the optical and gamma-ray emission, we analyse data from the \textit{Fermi} Large Area Telescope (LAT) \citep{atwood, ackermann12} for 2017 January 1 as well as over a longer time period covering the MAGIC observation window.
The LAT is an imaging high-energy gamma-ray telescope on board the \textit{Fermi} satellite, covering the energy range from about 20\,MeV to more than 300\,GeV. Its field of view covers about 20\% of the sky at any time and, when working in survey mode, covers the whole sky every three hours.\\
The data are reduced and analysed using \texttt{fermipy}\footnote{http://fermipy.readthedocs.io/en/latest/} \citep{mwood17} with the latest release of the Pass 8 Fermi Science Tools \footnote{https://fermi.gsfc.nasa.gov/ssc/data/analysis/documentation/}. We use the instrument response functions (IRFs) \texttt{P8R2\_SOURCE\_V6}, the isotropic diffuse background template \texttt{iso\_P8R2\_SOURCE\_V6\_v06} and the galactic diffuse background emission model \texttt{gll\_iem\_v06} \citep{acero2016}.\\
We select all the photons in a region of interest (ROI) of radius $10^{\circ}$ around the coordinates of NGC\,1275, and perform a binned likelihood analysis using three bins per energy decade in an energy range from 100\,MeV to 10\,GeV for the spectrum evaluation and the energy range 0.1--300 GeV for the light curve. All the 3FGL (third Fermi Gamma-ray LAT catalog, \citet{3fgl}) sources within $15^{\circ}$ from the source position are included in our model, along with the galactic and isotropic diffuse emission.

\subsubsection{\textit{Fermi}-LAT light curve analysis}
The light curve analysis is performed using \textit{Fermi}-LAT data encompassing MAGIC and KVA observation windows. The data from MJD 57619.5 to MJD 57820.5 are divided in 24 hours bins (bin center at midnight), and in each time bin a likelihood analysis is performed. The normalizations of all the sources within a radius of 5$^\circ$ from the source position are let free to vary, while the spectral indexes are fixed to the catalog value. The normalizations of the diffuse components are kept fixed. NGC\,1275 spectrum is modeled with a simple power-law. The resulting integrated flux in an energy range of 0.1\,GeV to 300.0\,GeV is shown in the middle panel of Fig.~\ref{Fig:LC}. All the time bins in which the Likelihood fit returned a TS$<$25 are represented as upper limits.
The mean flux was estimated to be $(5.8\pm2.5)\times10^{-7}$\,cm$^{-2}$\,s$^{-1}$. We fit the light curve with a constant function in the time range from MJD 57620.0 to 57820.0. This yield a $\chi^2$/d.o.f. of $358.2/151$ without considering the upper limits. 

\subsubsection{\textit{Fermi}-LAT spectral analysis}
Given the short observation time of MAGIC (few hours per night) simultaneity of data for a spectral analysis is limited by the minimal exposure time necessary to get a reliable spectral analysis of the LAT data; considering the short variability observed of $\sim$ 10 h, for the spectral analysis of 2017 January 1 we selected a time span of 12 hours, centered around midnight: MJD\,57753.75--57754.25. All the normalizations of sources within a radius of 5$^\circ$ from NGC\,1275 position are left free in the fit while the spectral indexes are fixed to the catalog value. The normalizations of the diffuse components is also fixed given the difficulty to fit their contribution in such a small integration time. NGC\,1275 is modeled with a \texttt{PowerLaw2}:
${\rm d}F/{\rm d}E=(F_0 (\Gamma + 1) E^{-\Gamma})/(E^{\Gamma+1}_{\rm max} - E^{\Gamma+1}_{\rm min})$, allowing the errors on the integrated flux ($F_0$) to be evaluated directly by likelihood. Converting to a simple \texttt{PowerLaw} form: $\mathrm{d}F/\mathrm{d}E=f_0(E/E_0)^{-\Gamma}$, and evaluating $E_0$ as the point at which $f_0$ and $\Gamma$ show the minimum correlation (\textit{decorrelation energy}), instead of arbitrarily fixing it, we obtain:
$f_0 = (7.03 \pm 1.26) \times 10^{-10}$\,MeV$^{-1}$\,cm$^{-2}$\,s$^{-1}$, $\Gamma=1.80 \pm 0.17$, $E_0 = 0.56$\,GeV. \newline In the 12 h dataset the source shows a test statistic (TS) of 55.83 (determined from the likelihood ratio of the source / no source hypothesis), showing a significance of $\sim$7.5$\sigma$.

\subsection{Combined \textit{Fermi}-LAT and MAGIC spectral analysis for 2017 January 1}
\label{sec:joint-fit}
\begin{figure}
   \centering
   \includegraphics[scale=0.4]{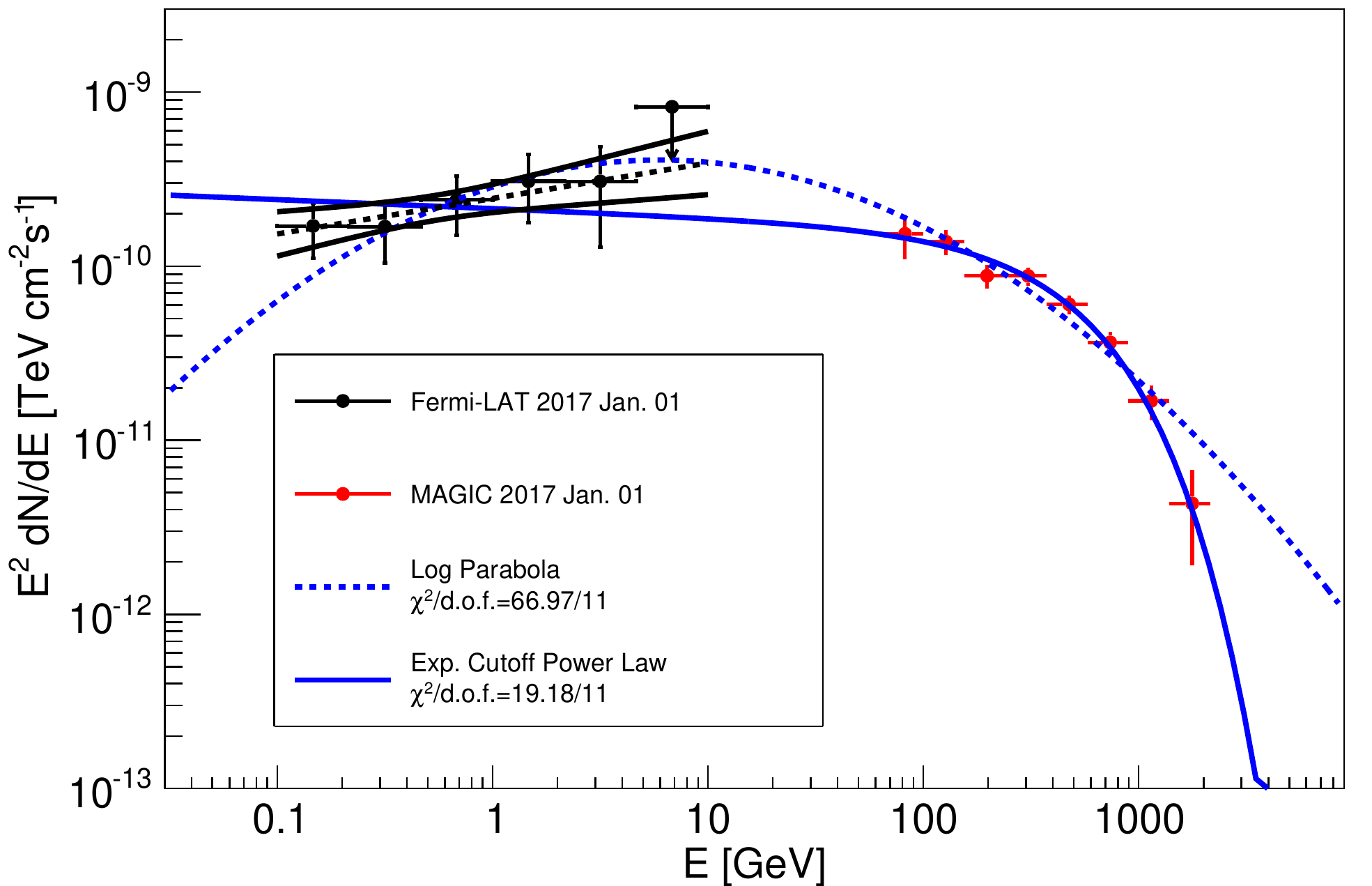}
   \caption{Combined spectral energy distribution from NGC\,1275 measured with \textit{Fermi}-LAT (low energy points and butterfly) and MAGIC (higher energy points) on 2017 January 01. The dashed and solid lines are fits to to the data points with a log-parabola and a power-law with exponential cutoff function, respectively. The fit results are given in the legend. Note that for the fit VHE bins in estimated energy, here not represented, are used. Unfolded spectral points (in true energy bins) are plotted for comparison.}
   \label{Fig:Spec_Fermi}%
\end{figure} 
A method for a joint spectral analysis of \textit{Fermi}-LAT and MAGIC data is described in \citet{moralejo17}. The spectral parameters are obtained maximizing a Poissonian likelihood describing the observed number of VHE events in the sky region around the source, and in three close-by background control regions, in each energy bin. The rate of excess events per (estimated) energy bin is then folded with the IRFs obtained from Monte Carlo simulations to obtain the expected flux. The \textit{Fermi}-LAT information is used in the following way to constrain the fit: assuming that the \textit{Fermi} data are fitted with a simple power-law, two additional terms are introduced in the Likelihood to anchor the high energy (HE) spectrum to the VHE one: $[(f - f_{\rm Fermi})/\Delta f_{\rm Fermi}]^2$, $[(\Gamma - \Gamma_{\rm Fermi})/\Delta \Gamma_{\rm Fermi}]^2$. Where $f_{\rm Fermi} \pm \Delta f_{\rm Fermi}$ and $\Gamma_{\rm Fermi} \pm \Delta \Gamma_{\rm Fermi}$ are the outcome of the \textit{\rm Fermi} power-law based spectral analysis and $f$ and $\Gamma$ are normalization and spectral index at the decorrelation energy. The allowed functions for the fit are always nested models of a simple power-law. The results of the joint fit follow in \mbox{Table~\ref{table:tab_spec_joint}} and are shown in Fig.~\ref{Fig:Spec_Fermi}. \newline
The joint analysis fit with \textit{Fermi}-LAT data confirms that for the brightest flare a power-law with exponential cutoff is preferred for fitting the spectrum. The position of the cutoff agrees with what estimated only using MAGIC data.

\begin{table}
\tiny    
\caption{Parameters of the joint spectral fit to the gamma-ray SED obtained with MAGIC and {Fermi}-LAT data. A power-law with exponential cutoff (EPWL) following 
${\rm d}N/{\rm d}E = f_0 (E/E_0)^{- \Gamma} \exp(-E/E_c)$ 
and a log parabola (LP) 
${\rm d}N/{\rm d}E = f_0 (E/E_0)^{- \Gamma - \beta \log(E/E_0)}$ 
are tested. The flux normalizations $f_0$ are given in units of $10^{-10}$\,TeV$^{-1}$\,cm$^{-2}$\,s$^{-1}$, and energies in GeV. Only statistical errors are given. The pivotal energy $E_0$ is evaluated to minimize the correlation between $f_0$ and $\Gamma$.}

\begin{center}
\begin{tabular}{c c c}     
\hline\hline
Fit  & LP & EPWL  \\\hline
$f_0$ & $34.2 \pm 1.1$  & $41.7 \pm 2.2$ \\
$\Gamma$ & $-2.76 \pm 0.03$ & $-2.05 \pm 0.03$ \\
$E_0$ & 180.77 & 198.21 \\
$\beta$ & $0.26\pm0.02$ & -  \\
$E_c$ & - & $492 \pm 35$ \\
$\chi^2$/d.o.f. & 66.97/11 & 19.18/11 \\
\hline
\end{tabular}
\end{center}
\label{table:tab_spec_joint} 
\end{table}

\subsection{KVA}

In the optical, NGC\,1275 is being monitored within the Tuorla blazar monitoring program \footnote{\url{http://users.utu.fi/kani/1m}} since 2009 October. In this work we discuss the observations in the R-band (640\,nm) performed with the Kungliga Vetenskaps Akademien (KVA) 35\,cm telescope in the time period covering the MAGIC observations.  
For the data reduction, the standard analysis pipeline is applied. Magnitudes are measured with differential photometry with an aperture of 5.0\(^{\prime\prime}\) and comparison stars from \citet{fiorucci1998}.\\
The resulting flux densities corrected for the host galaxy and the galactic extinction \citep{Schlafly2011} are shown in the bottom panel of Fig.~\ref{Fig:LC}. To study the flux variability, we fit the light curve from KVA in the R-band between 2016 September and 2017 February with a constant function and obtain a $\chi^2$/d.o.f. of $638.5/34$, thus implying significant flux variability. The mean flux during the period presented in this work is calculated to be $(9.5\pm1.3)$\,mJy. This is 1.6 times higher than the mean flux found during observations between 2009 October and 2011 February \citep{aleksic14c}. 

\subsection{Optical-Gamma-ray correlation analysis}

\begin{figure*}
   \centering
   \includegraphics[width=8cm]{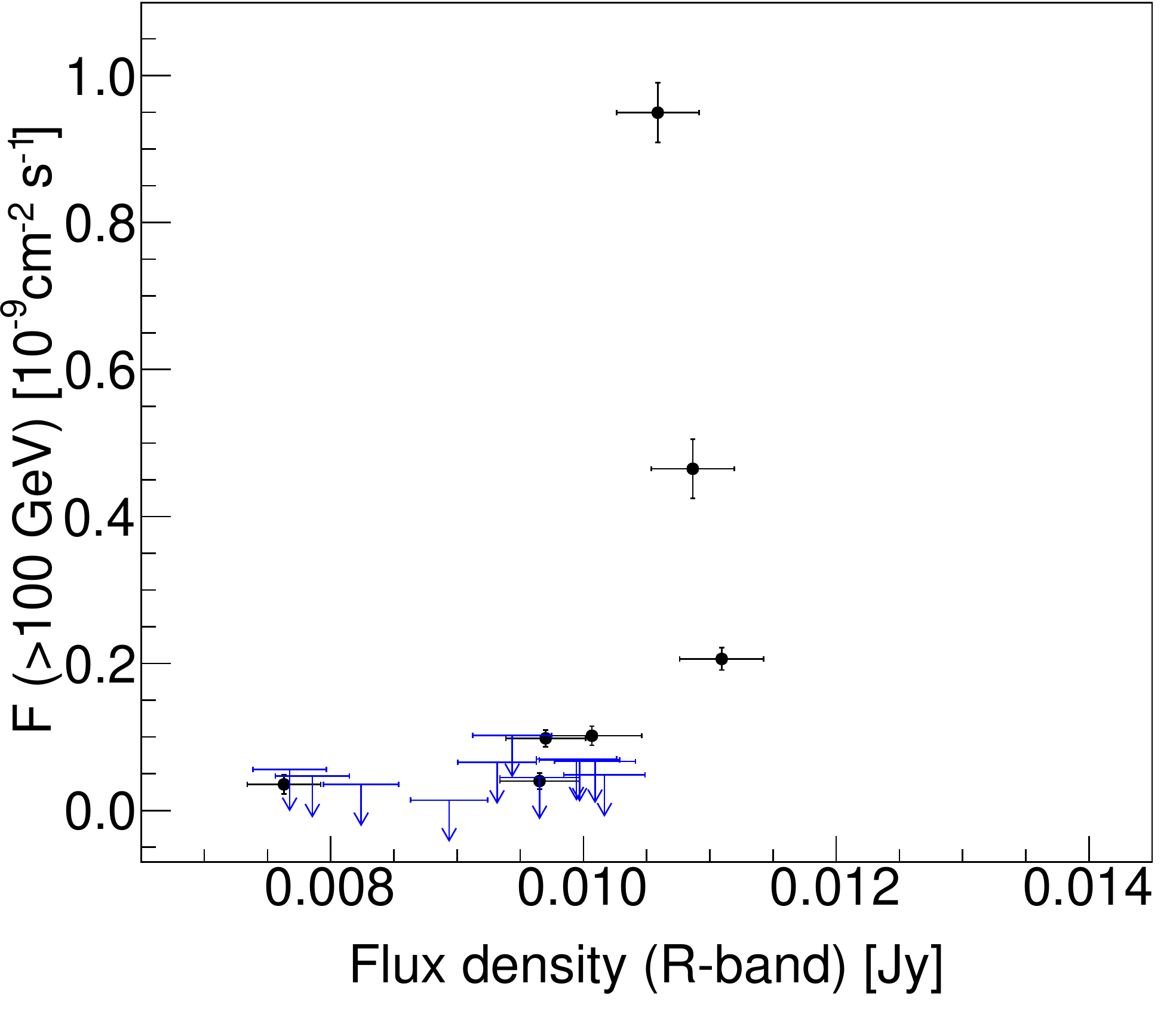}
   \includegraphics[width=8cm]{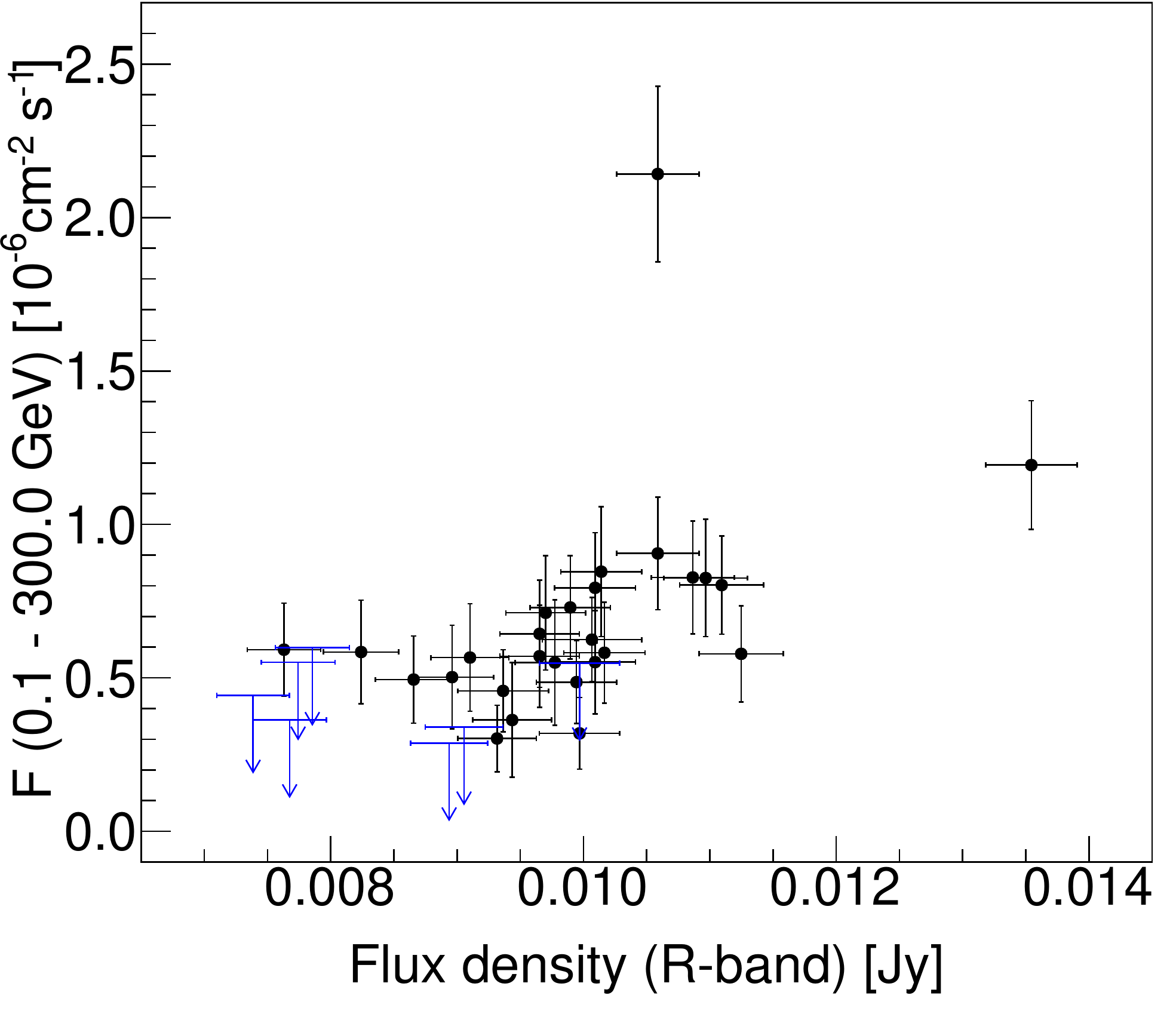}
   \caption{Gamma-ray flux versus optical flux density. Only measurements performed within less then a half of a day are included. \textit{Left panel:} MAGIC flux above 100\,GeV versus KVA R-band flux density. \textit{Right panel:} \textit{Fermi}-LAT flux between 0.1\,GeV and 300\,GeV versus KVA R-band flux density.}
              \label{Fig:Corr}%
\end{figure*}

We further investigate the correlation between the optical and the gamma-ray emission during the observation period. 
In Fig.~\ref{Fig:Corr} we plot the daily gamma-ray fluxes obtained by MAGIC and LAT measurements versus the flux density measurements in the R-band from KVA for observations performed within MJD $\pm$0.5. The linear correlation coefficient (Bravais-Pearson coefficient) is calculated to be 0.50 and 0.52 for MAGIC-KVA and LAT-KVA, respectively (upper limits are not used for the coefficient calculation). Thus, the optical emission is much less correlated with the gamma-ray flux with respect to the value of 0.79 (LAT-KVA) reported in \citet{aleksic14c}. Without the flaring nights (2017 January 1 to 2017 January 3), the coefficient increases to 0.82 and 0.72 for MAGIC-KVA and LAT-KVA. Similar to the study performed in \citet{aleksic14c}, we also fit the correlation plots with a linear ($F_\mathrm{gamma}=a \times F_\mathrm{opt}$) and a quadratic function ($F_\gamma=a \times F_\mathrm{opt}^2$). In a standard SSC flaring scenario, quadratic and linear correlation are expected between the optical and gamma rays, depending on the physical parameter that is causing the flare, see discussion in \citet{aleksic14c}. For LAT-KVA we find a $\chi^2 / {\rm d.o.f} =  56.67/27$ for the linear correlation and $\chi^2 / {\rm d.o.f} =  47.19/27$ for the quadratic. For MAGIC-KVA we find a $\chi^2 / {\rm d.o.f} =  557.22/6$ for the linear correlation and $\chi^2 / {\rm d.o.f} =  470.24/6$ for the quadratic. Thus no correlation between gamma-ray and optical fluxes is found.

\section{Discussion}
\subsection{Size of the emitting region}
The flux variability gives the possibility to estimate the size of the gamma-ray emitting region. Assuming a spherical region, we can estimate its radius $R$ using causality arguments. 
The radius of the emission source at redshift $z$ is constrained by the variability time scale $\tau_\mathrm{var}$ and can be estimated with $R\leq c\tau_\mathrm{var}\delta(1+z)^{-1}$, where $\delta$ is the Doppler factor to describe the relativistic beaming. The shortest flux-doubling time-scale found in this work is 611\,min. This correspond to a size of a spherical emission region of $R=\delta\cdot1.1\times10^{15}$\,cm. 
\citet{wilman2005} and \citet{scharwachter2013} measured a mass of the central black hole of NGC\,1275 of $M_{BH}=3.4\times10^8\,\mathrm{M}_\odot$ and $M_{BH}=8^{+7}_{-2}\times10^8\,\mathrm{M}_\odot$, respectively.  These masses correspond to a gravitational radius of $R_\mathrm{G}=GM/c^2=5.0\times10^{13}$\,cm and $1.2\times10^{14}$\,cm, respectively, and a light crossing time of $T_\mathrm{G}=R_\mathrm{G}/c=27.9$\,min and $65.6$\,min, respectively. Thus, the observed variability time scale is much larger than the event horizon light crossing-time but indicate an emission region one or two orders of magnitude smaller than the one proposed in \citet{aleksic14c} or \citet{Tavecchio2014}. \\
It is therefore necessary to investigate if this small emission region is transparent for TeV photons to escape. Taking into account that 1\,TeV photons were observed (see Fig.~\ref{Fig:Spec}), we calculate the Doppler factor which is necessary to avoid internal absorption via $\gamma\gamma$-pair production. 
Following Eq.~9 in \citet{abdo2011} for the optical depth for pair-production, we can establish the following condition in order to observe TeV photons
\begin{equation}
 \tau_{\gamma\gamma}\sim\frac{\sigma_\mathrm{T}D_\mathrm{L}^2F_0\epsilon_\gamma(1+z)}{10Rm_e^2c^5\delta^5}<1
\end{equation}
where $\sigma_T$ is the Thomson cross section, $D_L$ is the luminosity distance, $F_0$ the assumed peak of the target SED, $\epsilon_{\gamma}$ the energy of the hitting photon (1 TeV in this case), $z$ the redshift of the source, $R$ is the size of the emitting region, $m_e$ the mass of the electron and $c$ the speed of light. A simultaneous flux measurement in the $\sim$eV energy band (where the absorption with the TeV photons is supposed to happen) is available within the KVA dataset. For the night 2017 January 01 (MJD$_{\rm KVA}$ = 57753.94) a flux of $F_{\rm KVA} = 4.96 \times 10^{-11}$\,erg\,cm$^{-2}$\,s$^{-1}$ is observed, yielding a Doppler factor of $\delta_{\rm KVA} > 4.7$. This measurement, although corrected for the host-galaxy contribution, could be related to a region different than the one causing this particular VHE flare, as the lack of gamma-optical correlation is suggesting. We try then to provide another lower limit for the Doppler factor assuming a leptonic radiative model. From theoretical considerations we know that even in the most extreme flares the Compton dominance (i.e. the ratio between inverse Compton and the synchrotron peak) can not be greater than two orders of magnitude \citep{zacharias12, ghisetav10} and we use this constraint to derive $F_0$ as the peak of the synchrotron component. The SED Compton peak measured in section ~\ref{sec:joint-fit} is $6.42\times10^{-10}$\,erg\,cm$^{-2}$\,s$^{-1}$ (note that the log-parabola is used for this estimate, lacking the power-law with exponential cutoff SED of a local maximum, spectral index < -2). We get $\delta_{\rm SSC\,scenario} > 3.3$, with the conservative assumption $F_0 = F_{\rm Compton\;peak} / 100$. Using the afore mentioned Doppler factor values, we find the viewing angle $\theta_{\rm KVA} < 17^\circ$ and  $\theta_{\rm SSC\,scenario} < 12^\circ$, for any Lorentz factor $\Gamma_\mathrm{bulk}$. These are smaller than the results from radio observations, reporting $\theta>30^\circ$. For a large $\Gamma_\mathrm{bulk}$, a large viewing angle may even lead to a de-boosting. \citet{Hovatta2009} found a very small Doppler factor of 0.3 for NGC\,1275 based on the observation of the variability brightness temperature in the radio band while $\delta$=2 and 4 was assumed in \citet{aleksic14a} to model the source with a SSC scenario, these values are difficult to accommodate within the constraints found above.

\subsection{Physical models}
Since the lack of gamma-optical radiation correlation and the transparency constraint given by the small emission region hamper the use of a classical SSC leptonic scenario, we consider the feasibility of the following alternatives.
\subsubsection{Spine-layer model}
First we examine the spine-layer model in \citet{Tavecchio2014}. This model is based on a structured jet hypothesis and has been suggested to fit the broad-band emission of NGC\,1275. The fast spine with Lorentz factor $\Gamma_\mathrm{spine}=10-20$ in the inner part of the cylindrical jet is surrounded by a slower layer with  $\Gamma_\mathrm{layer}=2-4$. The low energy emission (radio to X-ray) in the broad-band SED should be dominated by the spine, whereas the high-energy emission should be predominantly produced by the layer.
An evidence of the structural configuration is given by the detection of a limb-brightened structure of the inner parsec-scale jet in high-resolution radio data reported by \citet{nagai2014}. \citet{Tavecchio2014} claimed that assuming $\theta\gtrsim25^\circ$ is incompatible with a scenario involving internal $\gamma\gamma$-pair production. Indeed, in this work we clearly detected photons $>1$\,TeV for which the optical depth becomes $\gg 1$ in this model since the requirement of a slow $\Gamma_\mathrm{layer}$ for a large $\theta$ leads to insufficient Doppler boosting. Thus, the spine-layer model from \citet{Tavecchio2014} it is not suitable to explain our data.\\
In what follows we study then the scenarios suggesting fast variability outlined in \citet{aharonian2017}, namely the magnetospheric model, the mini-jets model and the cloud-jet interaction. 

\subsubsection{Magnetospheric model}

As first described in \citet{BZ77}, a spinning black hole embedded in an external magnetic field can build up a force-free magnetosphere ($\mathbf{ E \cdot B = 0}$ along the magnetic field lines), supported by an electron-positron plasma generated by pair cascading. The injection of charges in the magnetosphere is commonly explained with the pair cascading of self-annihilating MeV photons produced from a radiatively inefficient accretion flow (RIAF) via free-free emission. During phases of low accretion the minimal charge density $n_{GJ} = \mathbf{\Omega_F} \cdot \mathbf{B} / (2 \pi e c)$ (where $\mathbf{\Omega_F}$ is the angular frequency of the dragged magnetic field lines), required to keep the magnetosphere force-free, could not be sustained. Charge-starved region (gaps) with a significant component of $\mathbf{E}$ parallel to $\mathbf{B}$ would then arise as electrostatic accelerators for the leptons. Curvature and Inverse Compton (IC) photons generated by the leptons within the gap and synchrotron and IC photons from the leptons cascaded outside will made the electromagnetic radiation of the source. This model has already been applied to radio galaxies, in particular in \citet{neronov07} and \citet{levinson2011} to M87 and Sgr A* and in \citet{hirotani16a} to IC 310. \newline
Since the gap can release only a part of the electromagnetic power extracted from the BH, the Blandford-Znajek (BZ) mechanism poses an upper limit to the gap luminosity
\begin{equation}
L_{\rm BZ} = 10^{21}\,a_\text{*}^2\,M^2_{1}\,B^2 \mathrm{erg\, s^{-1}}
\label{eq:L_BZ}
\end{equation}
where $M_1 = M_{\rm BH} / (10 M_\odot)$, $a_{\text{*}}^{} \equiv J / (GM^{2}c^{-1})$ denotes the dimensionless BH spin parameter ($J$ its angular momentum), $B$ the intensity of the magnetic field threading the BH. Assuming that in the RIAF the equipartition magnetic pressure is half the gas pressure \citep{mahadevan97, levinson2011} allows to set
$B_{\rm eq}\approx4 \times 10^8\,\dot{m}^{1/2}\,M^{-1/2}_{1}\,\mathrm{G}$, where $\dot{m}$ is the accretion rate measured in units of the Eddington rate: $\dot{m} = \dot{M} / \dot{M}_{\rm Edd} = \dot{M} / (L_{\rm Edd} / \eta\,c^2)$ and $\eta$ is the efficiency of conversion from mass to radiant energy, typically assumed $\approx 0.1$. Plugging this magnetic field value in \mbox{Eq.~\ref{eq:L_BZ}} produces
\begin{equation}
L_{\rm BZ} = 1.7\times 10^{38}\,a_\text{*}^2\,\dot{m}\,M_1 \mathrm{erg\, s^{-1}}.
\label{eq:L_BZ_max}
\end{equation}
To create a gap, the electron-positron density $n_{\pm}$ of the pairs cascaded by the MeV RIAF photons has to become less than $n_{\rm GJ}$. As shown in \citep{levinson2011} and remarked in \citet{hirotani16b} (Eq. 8) the ratio of these two charge densities is only dependent on $\dot{m}$ and $M_{\rm BH}$. The condition for a gap to be open $n_{\pm}/n_{\rm GJ}<1$ yields $\dot{m} < 3.1 \times 10^{-3}\,M_1^{-1/7}$, that substituted in \mbox{Eq.~\ref{eq:L_BZ_max}} returns, for our case
\begin{equation}
L_{\rm BZ} = 5.2\times 10^{35}\,a_\text{*}^2\,M_1^{6/7} \mathrm{erg\, s^{-1}} = 1.2 \times 10^{42} \mathrm{erg\, s^{-1}}
\label{eq:L_BZ_NGC1275}
\end{equation}
using $M_{\rm BH} = 3.4 \times 10^{8}\,M_{\odot}$ and assuming $a_{\text{*}}=0.9$.
Therefore the gamma-ray luminosity measured in the highest flux night overcomes by $\sim$ 3 orders of magnitudes the upper limit imposed by the BZ total power.
To obtain a more precise upper limit on the maximum gamma-ray luminosity we can examine \mbox{Fig. 25} of \citet{hirotani16b} that depicts the gamma luminosity for curvature and IC processes in a BH with mass $10^{9} M_{\odot}$ (same order of magnitude of NGC\,1275 BH mass estimated both in \citet{wilman2005} and {scharwachter2013}). In the ballpark of this BH mass we see that the minimum accretion rate needed to sustain pair production in the gap is $\dot{m}_{\mathrm{low}} \approx 6 \times 10^{-7}$ for which a maximum gamma-luminosity (IC dominated) of $\sim 3 \times 10^{40} \mathrm{erg\,s^{-1}}$ can be attained. \newline
The constraint on the total luminosity (both the BZ upper limit and the power actually radiated via curvature and IC processes) is strongly dependent on the assumption that the magnetic field of the RIAF is at the equipartition (simplification from \mbox{Eq. ~\ref{eq:L_BZ}} to \mbox{Eq. ~\ref{eq:L_BZ_max}}). To explain the huge gamma-ray luminosity in the IC310 flare detected in \citet{aleksic14b} ($L_{\gamma} \sim 2 \times 10^{44} \mathrm{erg\, s^{-1}}$) and its 3 orders of magnitude overcoming the allowed BZ power ($L_{\rm BZ} = 5.3 \times 10^{41} \mathrm{erg\, s^{-1}}$) \citet{hirotani16a} contemplated, for an extremely rotating BH ($a_\text{*} \geq 0.998$), an enhancement of magnetic field due to compilation of plasma near the BH horizon. To accommodate the measured $L_\gamma\approx10^{45}$\,erg\,s$^{-1}$ the BZ limit has to be increased, eventually overcoming the jet power $L_{\mathrm{BZ}} > L_{\mathrm{jet}}$, estimated to be $\sim 10^{44} \, \mathrm{erg}\,\mathrm{s}^{-1}$ for NGC\,1275 (see next paragraph for more details). Such an increase of magnetic field should be sustainable only on timescales smaller than the jet propagation timescale, i.e. the gap could be opened with a small duty cycle. 
The gamma-ray luminosity we report in this paper can be framed in a magnetospheric scenario sustained by a RIAF only in the hypothesis of an enhancement of the disk magnetic field in the proximity of the BH horizon from its equipartition value. This would imply allowing for the BZ luminosity a value larger than the jet power, sustainable only within a small duty cycle, as in the event of a flare. As remarked in \citet{hirotani16a} a complete numerical simulation is needed to investigate the possibility of such an enhancement of B near the black hole horizon.

\subsubsection{Mini-jets model}
In the mini-jets model \citep{giannios09, giannios10} it is assumed that the main jet with Lorentz factor $\Gamma_\mathrm{bulk}$ contains several mini-jets with $\Gamma_\mathrm{co}$ which are produced, e.g., by dissipation of magnetic energy in strongly magnetized plasma regions. Their relative motion with respect to the main jet results in a higher emitted Lorentz factor which can solve the opacity problem occurring when fast VHE variability is observed. In case of NGC\,1275 we may see the emission of the mini-jets pointing outside the jet cone. 
The lower limit of the jet luminosity required for the mini-jets scenario can be calculated with Eq.~37 in \citet{aharonian2017}:
\begin{equation}
 L_{\mathrm{j,jj}}>0.006\Phi\left(1+\alpha^2\right)^4(\Gamma_\mathrm{bulk}/10)^{-2}L_\gamma\xi^{-1},
\end{equation}
where $L_\gamma$ is the luminosity in gamma rays, $\mathrm{\xi=1}$ accounts for the conversion efficiency from the jet material to the outflow, and from the outflow to radiation; $\Gamma_\mathrm{bulk}=10$, and $\alpha=\theta\cdot\Gamma_\mathrm{bulk}=2$ are the jet bulk Lorentz factor and the normalized viewing angle, respectively.
We assume a filling factor of the mini-jets inside the jet of $\Phi=0.1$ which corresponds to the total number of mini-jets during a flaring event, the duty cycle of flares, their duration, and the variability time scale. With the parameters given above, we derive a minimum jet luminosity of  $L_{\mathrm{j,jj}}=4.1\times10^{44}$\,erg\,s$^{-1}$. 
In \citet{abdo2009a}, a jet luminosity of $L_\mathrm{j}\sim(0.6-4.9)\times10^{44}$\,erg\,s$^{-1}$ was found for NGC\,1275 based on the modeling of the broad-band SED with a single-zone synchrotron self-Compton model assuming either one proton per radiating electron or a ten times higher energy density of the protons than the electrons. Using a different method, \citet{dunn2004} found a total power of $L_\mathrm{j}=(0.3-1.3)\times10^{44}$\,erg\,s$^{-1}$ required to inflate the radio lobes of NGC\,1275 against the pressure of the hot cluster gas. The two numbers are likely compatible given the entrainment (work against the interstellar and intergalactic medium). Thus, the mini-jets model can account for the observed gamma-ray emission assuming the range of the jet luminosity inferred by \citet{abdo2009a} but has difficulties to explain the measurements in case of smaller jet power of NGC\,1275. The situation further worsen for a higher filling factor. 

\subsubsection{Cloud in jet model}
In the cloud-jet interaction model \citep{barkov12, barkov12b} a VHE flare is explained by an obstacle moving through the jet or vice versa. For example, such an obstactle can be a star with a high mass-loss rate causing the formation of a cloud out of the lost material due to pressure in the jet. Interactions of colliding protons at a bow shock located at the jet-cloud interface produce a single peak in a VHE light curve.   
For this scenario, the lower limit of the jet luminosity can be estimated with Eq.~43 in \citet{aharonian2017}:
\begin{equation}
 L_{\mathrm{j,cj}}>0.025\left(1+\alpha^2\right)^4(\Gamma_\mathrm{bulk}/10)^{-2}L_\gamma\xi^{-1}.
\end{equation}
This gives a minimal jet power of $L_{\mathrm{j,cj}}=1.7\times10^{46}$\,erg\,s$^{-1}$ required to explain the NGC 1275 observations with a cloud-jet interaction model. This result clearly exceeds the inferred values for the jet power. \\
In the estimations presented a normalized viewing angle of $\alpha=2$ was assumed for NGC\,1275. The required jet powers in the mini-jets as well as in the cloud-jet interaction model would increase if a larger value for $\alpha$ is assumed, hence, making those scenarios more unlikely. 

\section{Conclusions}

In this work we present VHE gamma-ray data of NGC\,1275 measured in 2016 September to 2017 February with the MAGIC telescopes. We found several nights in which NGC\,1275 was in a high state with respect to the flux previously measured in the 2009--2011 campaigns. For the brightest flare around 2017 January 1 a value fifty times higher was measured, characterized by a flux-doubling times-scale of $\sim611$\,min which equals to 22 times the light crossing time at the black hole event horizon. The spectra from different flux states are generally harder than the ones from previous campaigns and can not be described with a simple power-law function. The combined spectral analysis of \textit{Fermi}-LAT and MAGIC data from 2017 January 1 yields good fit results when assuming a power-law function with exponential cutoff revealing a cutoff energy of $(492\pm35)$\,GeV. \newline
Furthermore, investigating the correlation of the optical and gamma-ray emission by comparing MAGIC, \textit{Fermi}-LAT and KVA (R-band) light curves, we find no correlated variability of the optical flux density around the time of the VHE flare.\\
Considering the observations in the lights of different emission models, the fast flux variability is constraining the size of the gamma-ray emission region to a value one or two orders of magnitude smaller than the one used within the SSC scenario proposed in \citep{aleksic14c} or the spine-layer in \citet{Tavecchio2014}. A higher Doppler factor than the one assumed in \citep{aleksic14c} would be needed to avoid absorption in the SSC scenario, implying a viewing angle in tension with the large value observed in radio. Absorption of the highest energy photons via $\gamma-\gamma$ pair production in such a small emitting region would be also dominant in a spine-sheath scenario, thus excluding this theoretical model standing the significant emission above 1 TeV measured. Among the alternative scenarios for fast variability presented in \citet{aharonian2017}, the mini-jets model and the cloud-jet interaction, probably fail because of a large jet power necessary to reach the observed gamma-ray luminosity. A hard limit on the maximum luminosity expected for a magnetospheric model can be estimated from the maximum extractable BZ power (under the assumption that the magnetic field in the disc is at the equipartition value) and from the condition on the accretion rate in Eddington units $\dot{m}$ needed to open a gap (Eq. 8 and 4 in \citet{levinson2011} and \citet{hirotani16b} respectively). The only possibility to fit the enormous (10$^{45}$\,erg\,s$^{-1}$) luminosity measured in this paper for the higher flaring state within the strong upper limit posed by the BZ power (10$^{42}$\,erg\,s$^{-1}$), as suggested in \citet{hirotani16a}, would be an enhancement of the magnetic field threading the BH horizon from its equipartition value, increasing the extractable BZ power even beyond the jet power. This increase has to happen on a time scale smaller with respect to the jet propagation timescale (e.g. during a flaring event) and yet has to be proved by numerical studies. The luminosities and the corresponding fast variability hereby reported pose a challenge to the actual models for fast variability of VHE gamma-ray emission in AGN.


\begin{acknowledgements}
We would like to thank the Instituto de Astrof\'{\i}sica de Canarias for the excellent working conditions at the Observatorio del Roque de los Muchachos in La Palma. The financial support of the German BMBF and MPG, the Italian INFN and INAF, the Swiss National Fund SNF, the ERDF under the Spanish MINECO (FPA2015-69818-P, FPA2012-36668, FPA2015-68378-P, FPA2015-69210-C6-2-R, FPA2015-69210-C6-4-R, FPA2015-69210-C6-6-R, AYA2015-71042-P, AYA2016-76012-C3-1-P, ESP2015-71662-C2-2-P, CSD2009-00064), and the Japanese JSPS and MEXT is gratefully acknowledged. This work was also supported by the Spanish Centro de Excelencia ``Severo Ochoa'' SEV-2012-0234 and SEV-2015-0548, and Unidad de Excelencia ``Mar\'{\i}a de Maeztu'' MDM-2014-0369, by the Croatian Science Foundation (HrZZ) Project IP-2016-06-9782 and the University of Rijeka Project 13.12.1.3.02, by the DFG Collaborative Research Centers SFB823/C4 and SFB876/C3, the Polish National Research Centre grant UMO-2016/22/M/ST9/00382 and by the Brazilian MCTIC, CNPq and FAPERJ.\\
We would like to thank F. Rieger for useful discussions.
\end{acknowledgements}

%
%

\bibliographystyle{aa} 
\bibliography{references}
\end{document}